# Predicting Chemical Reaction Outcomes Based on Electron Movements Using Machine Learning

Shuan Chen[1]†, Kye Sung Park[1]†, Taewan Kim[2], Sunkyu Han[2] and Yousung Jung[1,3]*

1. School of Chemical and Biological Engineering, Institute of Chemical Processes, Seoul National University, Seoul, South Korea

2. Department of Chemistry, Korea Advanced Institute of Science and Technology, Daejeon, South Korea

3. Institute of Engineering Research, Seoul National University, Seoul, South Korea

†These authors contributed equally to this work.

*Corresponding author. Email: yousung.jung@snu.ac.kr

## Abstract

Accurately predicting chemical reaction outcomes and potential byproducts is a fundamental task of modern chemistry, enabling the efficient design of synthetic pathways and driving progress in chemical science. Reaction mechanism, which tracks electron movements during chemical reactions, is critical for understanding reaction kinetics and identifying unexpected products. Here, we present Reactron, the first electron-based machine learning model for general reaction prediction. Reactron integrates electron movement into its predictions, generating detailed arrow-pushing diagrams that elucidate each mechanistic step leading to product formation. We demonstrate the high predictive performance of Reactron over existing product-only models by a large-scale reaction outcome prediction benchmark, and the adaptability of the model to learn new reactivity upon providing a few examples. Furthermore, it explores combinatorial reaction spaces, uncovering novel reactivities beyond its training data. With robust performance in both in- and out-of-distribution predictions, Reactron embodies human-like reasoning in chemistry and opens new frontiers in reaction discovery and synthesis design.

## Main

In organic chemistry, a reaction mechanism is a theoretical trajectory that describes how the electron moves within organic molecules in a chemical reaction. For polar reaction, the arrow pushing diagram is one of the most widely-accepted ways to represent the reaction mechanism[1], where the electron flows from an electron-rich (source) lone pair or bond to an electron-poor (sink) atom or bond. By sequentially applying arrow-pushing diagrams to the reactants, intermediates are generated until the final products are formed, at which point electron flow ceases. While accurate rationalization of reaction mechanisms has great potential for advancing chemical science, proposing reaction mechanisms remains challenging even for experienced chemists due to the exponentially growing number of potential pathways. To this end, rule-based expert systems have been proposed to predict the reaction mechanism for specific reaction types[2,3]. However, these methods focus on a relatively narrow scope of reaction types and are difficult to scale up to predict a larger reaction space due to the mass number of manually curated rules and complicated rule dependencies.

Recently, chemists have devoted extensive efforts to develop machine learning models for reaction outcomes prediction[4–9] on diverse chemical reactions in the US Patent and Trade Office (USPTO) reaction dataset[10], yet none of these models are able to propose the actual reaction mechanism. Although Bradshaw et al.[11] proposed a model that generates the electron path of chemical reaction, the generated electron paths fail to match the realistic reaction mechanism due to the approximated mechanism label in the training data. The primary bottleneck in developing a reaction mechanism prediction model is the lack of reaction mechanism data. To address this bottleneck, a recent study by Joung et al.[12] developed a bottom-up method to label the reaction mechanism by iteratively applying elementary reaction templates until the reported products are found. Chen et al.[13] developed a mechanism labeling tool named MechFinder, which generates reaction mechanisms from a given chemical reaction based on expert knowledge, to label the arrow-pushing mechanisms of 31k reactions from the USPTO reaction dataset and validated them by expert chemists.

In this work, we expanded the reaction mechanism dataset from 31k to 213k reactions by expanding the reaction labeling scope of MechFinder (Fig. 1a) and developed a graph-based machine learning model, *Reactron*, to predict reaction outcomes by sequentially predicting the electron movement in the fundamental arrow-pushing diagram (Fig. 1b). Our experiment suggests Reactron has superior prediction accuracy compared to previous reaction outcome prediction models and the potential to explore unknown chemical reactions with detailed proposed mechanisms (Fig. 1c). By fine-tuning Reactron with a few reaction examples, we show its great flexibility and potential for predicting challenging named reactions such as rearrangement or functional group interconversion reactions from complex chemical structures, while the existing language-based model struggles to generate chemically valid predictions (Fig. 1d). Finally, we apply Reactron as a chemical reaction discovery engine by performing a large-scale combinatorial chemical reaction prediction that interpolates the learned reaction space, rediscovering multiple chemical reactions shown in literature published across the 1960s to the 2020s that do not exist in the training set (Fig. 1e).

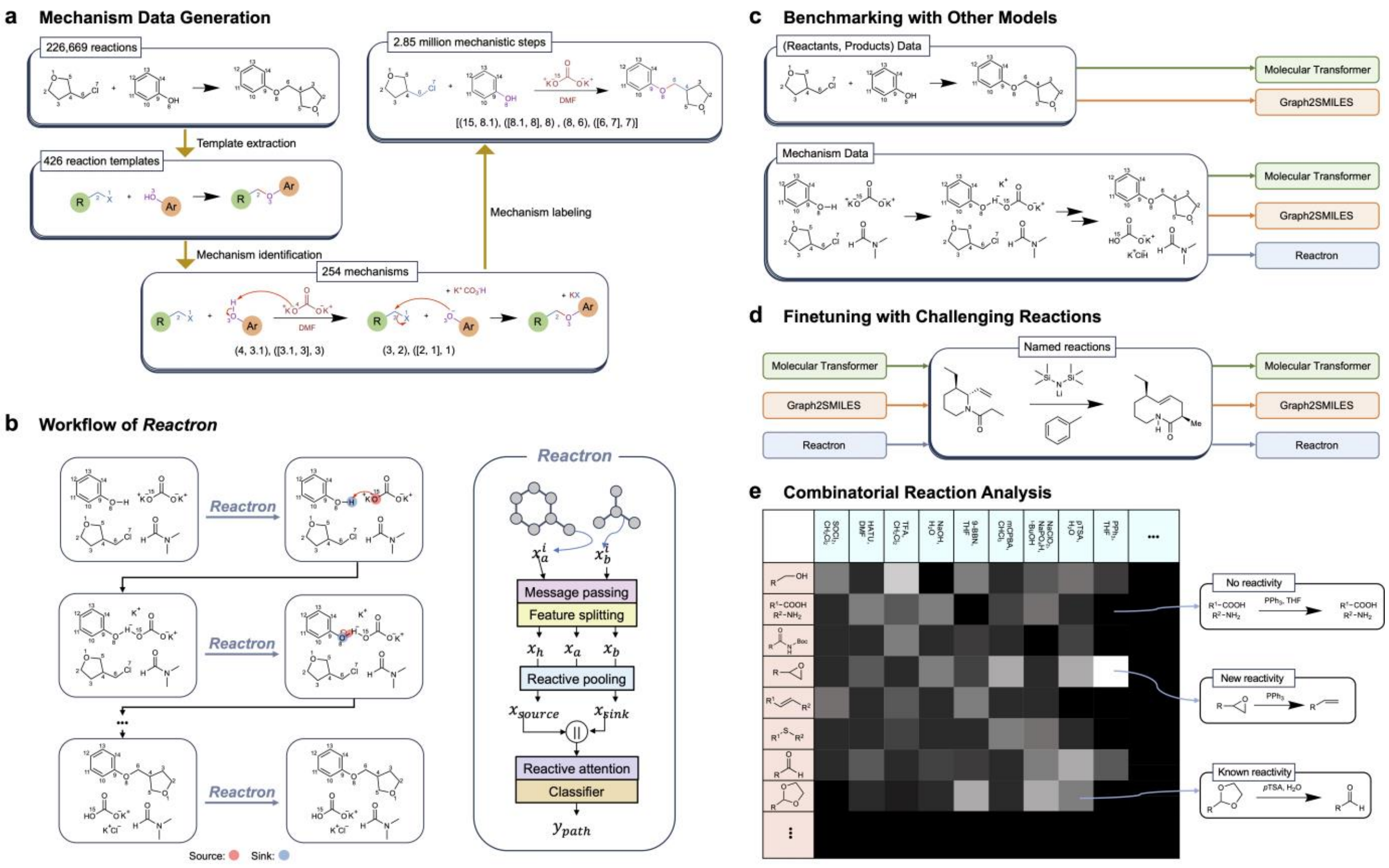

**Fig. 1:** The overall framework of this work. a, Reaction mechanism data preparation. b, The workflow and the architecture of the proposed machine learning model Reactron. c, Benchmarking Reactron with existing models on USPTO reaction dataset. d, Finetuning Reactron to for named reactions prediction. e, Interpolating chemical data for reaction discovery.

## Model development

To generate the training data for the reaction mechanism, we apply LocalMapper[14] to annotate the atom-mapping for each reaction and MechFinder[13] to generate reaction mechanisms from chemical reactions by analyzing the top-500 popular reaction templates and manually curated 254 mechanistic templates[15,16] from the USPTO-480k reaction dataset[17] (Fig. 1a). Among all the 480k reactions, we labeled 226,669 reactions after removing the organometallic or radical reactions, resulting approximately 2.85 million labeled mechanistic steps. The train-test split of this dataset follows the original USPTO-480k dataset, yielding 193,052 reactions in the training set, 14,272 reactions in the validation set, and 19,345 reactions in the test set. Within the train dataset, reactions belonging to the 100 least popular reaction templates were intentionally held out in order to examine the generalizability of our model. The test reactions belonging to these reaction templates are later called out-of-distribution (OOD) reactions in the later results analysis. More details about the data labeling process are given in Methods.

Next, we train a novel ML-based electron flow prediction model, Reactron, on the generated reaction mechanism dataset described above. We design Reactron to predict the source and sink of each arrow pushing occurring in each step of the reaction mechanism to simulate the arrow-pushing diagrams of electron pairs that occur in chemical reactions. By iteratively applying Reactron to the reactants, the intermediates can be obtained until no more reactivity is predicted in the given set of molecules, which is defined as the final product. An example of predicting the electron flow of an $S_N2$ reaction is shown in Fig. 1b. The molecule structures generated by the abovementioned process is translated into molecule graphs, where the atoms and bonds in the given structures are represented as the nodes and edges in the graphs. The features of heavy atoms $x_a$ and bonds $x_b$ in the molecule are initialized by their physical property, such as atom types or bond types. Subsequently, the source and sink candidates of given molecule graphs are predicted by Reactron with three major components: message passing[18], reactive pooling[8], and reaction path prediction. These three components are responsible for encoding the local chemical environment, filtering non-reactive species, and selecting the most plausible reaction path within all potential mechanism candidates, respectively. The overall architecture of Reactron is given in Fig. 1c. The full set of initial features for atoms and bonds can be found in Table S7, and the details about the model architecture and the mathematical formulas can be found in Methods.

## Benchmarking Reactron with previous models

To highlight the performance of Reactron, we use top-k prediction accuracy, which is the percentage of reactions that have the same SMILES string with the ground truth in the top-k predictions, for evaluation. To clearly show the different levels of performance in predicting the reaction outcome and mechanism, their prediction accuracies are compared separately. The accuracy of predicting the reaction outcome is described by product accuracy, and the accuracy of predicting the electron flow is described by mechanism accuracy. To assess the performance of Reactron compared to previous models, we trained two language-based reaction outcome prediction models, Transformer[6] and Graph2SMILES[9], as the strong baselines[12] for comparison. For each comparing model architecture, we trained two different models, one that only predicts the reaction products (-P models) and one that predicts all the reaction mechanisms (-M models). See the details of training the Transformer and Graph2SMILES models in Methods.

The results of top-k product and mechanism accuracy on the USPTO reaction dataset are shown in Fig. 2a and 2b. Reactron consistently outperforms Transformer and Graph2SMILES-based models, with 0.6% and 38.0% top-1 accuracy improvement compared to the second-best baselines for top-1 product and mechanism accuracy, respectively. Regarding the prediction speed, Reactron requires significantly less time, with nearly 50 times and 40 times faster than the other two mechanism prediction models, respectively, to predict the full mechanism sequence for a given set of reacting molecules (Fig. 2c). To examine the generalizability of Reactron, we calculate the prediction accuracy for out-of-distribution (OOD) reactions, whose reaction templates are not seen in the training set, separately at Fig. 2d and 2e. Compared to the accuracies computed from all test reactions, the top-1 accuracies of Reactron shows huge drops from 96.4% to 48.5% for product prediction and from 94.8% to 36.7% for mechanism prediction. Our analysis shows that Reactron consistently outperforms the baseline models in predicting both reaction products and mechanisms, regardless of the number of mechanistic steps or reaction templates in the training set (Extended Data Fig. 1). Representative comparisons between Transformer-M and Reactron are provided in Extended Data Fig. 2.

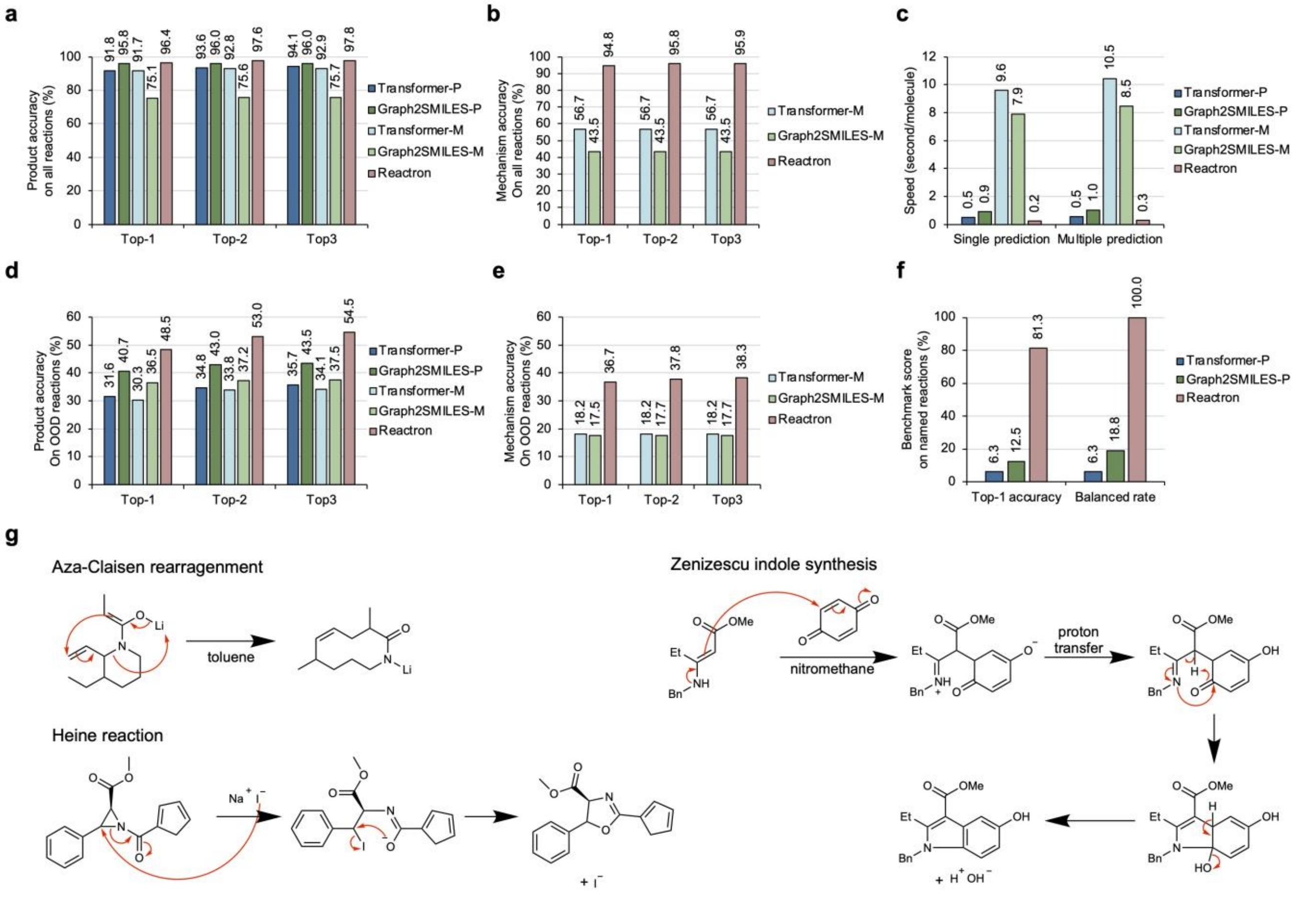

**Fig. 2:** The benchmark results of Reactron. a, The top-k product accuracies of Reactron compared with the baseline models on all reactions in the USPTO dataset. b, The top-k mechanism accuracies of Reactron compared with the baseline models on all reactions in the USPTO dataset. c, The prediction speed of each model for single-outcome and multi-outcome predictions on a single NVIDIA GeForce RTX 3090 GPU. d, The top-k product accuracies of Reactron compared with the baseline models on the out-of-distribution (OOD) reactions in the USPTO dataset. e, The top-k mechanism accuracies of Reactron compared with the baseline models on the out-of-distribution (OOD) in the USPTO dataset. f, The top-1 product accuracies and reaction validity of finetuned Reactron compared with the finetuned baseline models on name reactions. g, Examples of Reactron predictions on named reactions.

## Finetuning for named reactions

To further examine the capability of Reactron to predict reactions that are more challenging for experienced chemists, we manually collected 48 named reactions, including 16 distinct reaction types, from the textbook[19]. This additional reaction dataset consists of a wide range of reaction mechanisms, including different types of rearrangement reactions and complex reactions involving multiple mechanistic steps. We finetuned the Reactron using two reactions from each reaction type and test the finetuned model with the remaining one. Similarly, we finetuned the four comparing models to learn to predict the named reactions using the same data split. list of the sampled named reactions and the predictions can be found in Supplementary Information S4.

The results of finetuning Reactron and the baseline models, including the prediction accuracy and ratio of correct atom balance, are shown in Fig. 2f. After finetuning, Reactron accurately predicts 13 out of 16 test reactions, whereas the other models fail to predict most of the reactions. Notably, we found that baseline models fail to predict balanced reaction outcome in most of the cases (Supplementary Information S6), with extremely low ratio of correct atom balance, where the models eliminate or generate new atoms in the products that exist or do not exist in the input reactants. Although test reactions included complex mechanistic behaviors that are vastly different from those in the training set, Reactron demonstrated remarkable robustness by accurately predicting complex reactions after finetuning. It correctly identified mechanistic pathways for challenging reactions such as the Aza-Claisen rearrangement, Heine reaction, and Nenitzescu indole formations (Fig. 2g), which features nontrivial electron-flow behavior absent from the training data. This showcases the model's capacity to learn entirely novel pathways and diverse types of reactions only by exposing it to a few examples.

## Reaction discovery by combinatorial reaction prediction

In this section, we showcase Reactron as a tool to discover novel reactions by predicting virtually designed chemical reactions that are never seen in the training set. To generate a matrix of virtual reactions, we sample substrates from the chemical reactions that yield the same reaction template, and extract the reagents frequently used in the USPTO reactions (Methods). By combinatorial pairing these substrates and reagents, we can generate a large matrix that interpolate the unknown reaction space between the known reactions. In this study, we sampled 1,200 substrate sets (30 reactions from each reaction template) and extracted 250 unique reagent sets from the reactions corresponding to the top-400 most popular reaction templates by pairing each substrate with each of the unique reagent sets, a total number of 100,000 template-reagent pairs and 3,000,000 reaction entries were generated and their reactivities were predicted by Reactron. To fully demonstrate the reactivity prediction capacity, the Reactron model used in this experiment is trained on all the reactions in the USPTO dataset, including those in the train, validation, and test sets.

By extracting the reaction template of the predicted reactions, the predictions were classified into three classes: inconsistent or no reactivity, known transformation, and new transformation (Fig. 3a). Since most of the chemical reactions would not occur by simply matching random reagents to a set of reactants and Reactron shows low prediction accuracy for low-score predictions (Extended Data Fig. 3), we filter out 79.4% the non-reactive reactions or predictions with prediction scores lower than 0.5. Among the remaining reactions, 18,354 (18.4%) of the them were labeled as "known transformation" as the reactions give the reaction templates shown in the train set or the predicted electron movements are the subsets of the labeled mechanism in the train set. Finally, we narrowed down the analysis to 2,283 (2.3%) template-reagent pairs that show consistently novel transformation, including the reactions giving unseen reaction templates and the predicted electron movements do not belong to any subset of the labeled mechanism. More details about the definition of new transformation can be found in Methods.

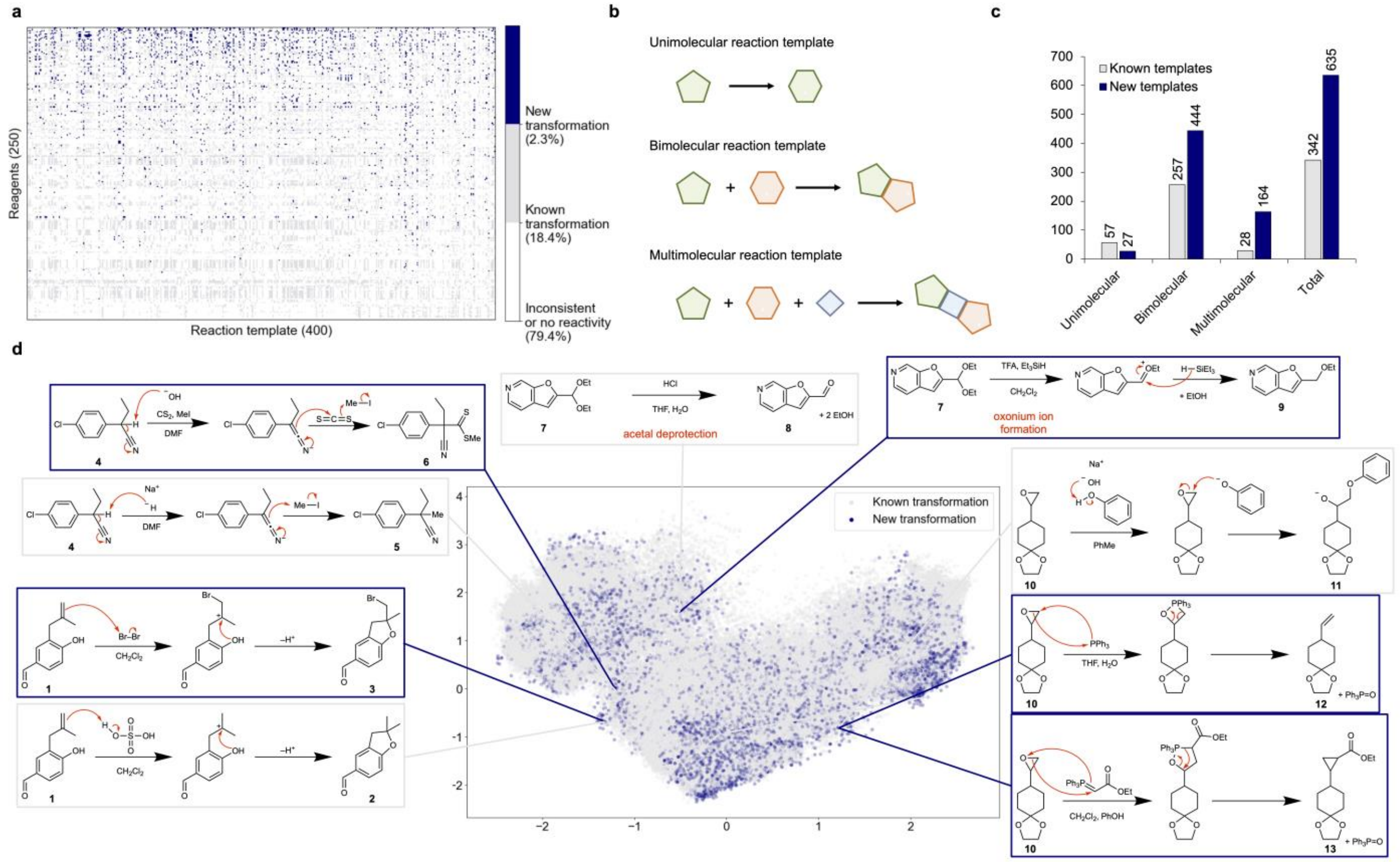

**Fig. 3:** Analysis of combinatorial reaction predictions. a, A reactivity matrix showing the predicted outcomes for various combinations of 400 reaction templates and 250 reagents, with color-coded indicators for new transformation (blue), known transformation (gray), and inconsistent or no reactivity (white). b, The cartoons of illustrating unimolecular, bimolecular, and multimolecular reaction templates defined in this study for reaction discovery analysis. c, The bar chart of analyzing the number of unimolecular, bimolecular, multimolecular, and all reaction templates within the known and new reaction templates. Note that the functional groups are not included in the reaction template here to more precisely capture the transformations distinct from the known ones. Therefore, the total number of known reaction templates here is 342 instead of 400. d, The principal component analysis (PCA) plot of the reactant space in our combinatorial reaction prediction and the visualization of pairwise comparison of known and new transformation discovered in this study. The examples of known transformations and new transformations are illustrated in the blue boxes and gray boxes, respectively.

To better understand the types of reactions discovered by our combinatorial reaction prediction experiment, we categorized the known reaction templates and new reaction templates into unimolecular, bimolecular, and multimolecular reaction templates (Fig. 3b). Within the newly discovered reaction templates, we found that 27 unimolecular reaction templates, 444 bimolecular reaction templates, and 164 multimolecular reaction templates are discovered in our experiment, which are 47%, 173%, and 586% compared to the known ones, respectively (Fig. 3c). It is not surprising that the number of discovered multimolecular reaction templates are much more than the ones recorded in the train set because of the diverse combinations of reactants and reagents paired in the experiment. Next, we visualized the predicted reactions, including the ones with known transformation and new transformation, by a principal component analysis (PCA) plot using the 2048-bit Morgan Fingerprints of substrates and reagents in each reaction, at Fig. 3d. While the majority of the reactions (88.2%) exhibit known transformations in the PCA plot, novel transformations could be discovered by interpolating between the known transformations using different reagents on the same substrates. We show five examples at the different locations at the PCA plot and compare them with the reactions in the train set using the same substrates but different reagents.

Alkene **1** (Fig. 3d), initially trained for intramolecular cyclization catalyzed by acid to yield ether **2**, underwent a similar transformation when bromine replaced the acid, resulting in bromide **3**. Notably, despite bromine being primarily trained for aryl bromination or α-bromination, the model accurately predicted the chemoselectivity between the alkene and aromatic group. Methylation of nitrile **4**, when modified with addition of carbon disulfide, correctly anticipated the formation of dithioate **6**, reflecting accurate reactivity sequence. Hydrolysis reaction of acetal **7** to give aldehyde **8**, was revisited with addition of triethylsilane and trifluoroacetic acid, yielding ether **9**. It proceeds through the formation of an oxonium ion intermediate, followed by reduction using triethylsilane[20]. Here, the model successfully identified triethylsilane as a compatible reducing agent, avoiding the use of stronger agents like lithium aluminum hydride, which could have led to adverse interactions with trifluoroacetic acid. This highlights the model's capacity to account for reagent compatibility and predict outcomes that align with practical synthetic considerations. For reactions involving epoxides, the model demonstrated its predictive strength in adapting trained mechanisms to new reagent combinations. For example, epoxide **10** was initially trained to undergo ring opening with a base and nucleophile, as shown with phenoxide attacking epoxide **10**. When the reagent was altered to triphenylphosphine, the model predicted olefination, yielding alkene **12**. To our surprise, the predicted electron flow matched with a recently published mechanistic proposal[21]. Another exploration revealed that epoxide **10** when treated with a Wittig reagent, underwent cyclopropanation to form compound **13**. The predicted electron flow aligns with the same mechanistic pathways originally suggested by a previous publication[22]. Overall, these results highlight Reactron's exceptional ability to uncover new reactivity and predict novel transformations with precision. Its capacity to adapt trained mechanisms to untested conditions establishes it as a powerful tool for advancing synthetic chemistry.

To further emphasize the capability of Reactron over the baseline models on discovering new reactions, we randomly sampled 100 reactions discovered in this experiment and applied the product prediction baseline models to predict their products. The results show that none of the models succeeded in reproducing any results we found in this experiment at top-1 predictions, where 46 and 47 predictions generated by Transformer-P and Graph2SMILES-P model, respectively, were found to have incorrect atom balance.

## Discussions

Through the comprehensive experiments. we show the superior prediction performance of Reactron on the benchmark dataset while providing detailed mechanistic explanations for how final products emerge from input molecules. By directly predicting electron movement to infer reaction outcomes, Reactron ensures 100% validity of its predictions, offering a robust and transparent (as opposed to a "black box") framework for interpreting chemical transformations. Moreover, its adaptability for learning new reactivity after exposed to few examples open the door of applying Reactron to quickly explore broader chemical reaction spaces. Our high-throughput combinatorial reaction prediction experiment demonstrates the great potential of using Reactron as a promising reaction discovering engine by interpolating the reaction space. Future work will focus on extending Reactron to discover reactions including radical- and organometallic-based transformations as suitable datasets become available.

## Methods

## Reaction Mechanism Data Curation

To generate the training data for the reaction mechanism, we apply MechFinder[13] to generate reaction mechanisms from chemical reactions by leveraging the reaction templates[15,16] and mechanistic templates. First, the reaction template is automatically extracted by comparing the local chemical environment of each atom in the reactants and products according to their atom mappings. Next, the mechanistic template, which is a manually labeled template that describes the reaction mechanism of the corresponding reaction templates, is mapped. Finally, the reaction mechanism of the reaction template is reversely mapped back to the input reaction. In the case of multi-step reactions and missing reagents, the input reaction is separated into two individual reactions, and the necessary reagents are added back to the input reaction. More details and examples for the mechanism labeling process can be found in the Supplementary Information S1.

In this paper, we started with the reaction dataset extracted from the US Patent and Trade Office (USPTO)[10], which includes approximately 1.7 million publicly available reactions and is a popular reaction dataset commonly used in the reaction prediction community. This dataset was further cleaned by Jin et al.[17] by removing duplicate and erroneous reactions, resulting to approximately 480 thousand reactions, also known as the USPTO-480k dataset. After removing catalyst-containing reactions and no-reagent reactions, 366,233 reactions were remained. Next, we apply LocalMapper[14] to obtain the atom-mappings of the remaining reactions, and extract the reaction templates. A reaction is defined as catalyst-containing if the reaction contains any of the following metals: Au, Ag, Co, Fe, Ni, Pd, Pt, Rh, Zn in this work.

We start the data curation from the reactions recorded in the USPTO-480k[17] dataset. After labeling the atom-mapping of each reactions by LocalMapper[14] we extracted 24,905 reaction templates from this reaction dataset. From these large amount of reaction templates, we investigated the reaction mechanism for the top-500 popular reaction templates, which cover 79.6% of the reactions in the dataset (Extended Data Fig. 4). During the investigation, we found 73 reaction templates (covering approximately 11.8% of the reactions) are either non-arrow-pushing reactions (e.g., organometallic or radical reactions) or wrongly mapped reactions. Therefore, of the 500 analyzed reaction templates, 426 of them were applied to the MechFinder labeling tool to label the reaction mechanism dataset.

To better understand the distribution of the reaction types used in this work, we categorize the reactions labeled in this dataset into 6 different reaction superclasses and analyze the number of mechanistic steps within each reaction category. The superclasses and the distribution of mechanistic steps and examples are displayed in Extended Data Fig. 5. These reaction superclasses were first categorized by Claude Sonnet 3.5 by providing the mechanistic classes and further refined by human chemists. We found the majority of the labeled reactions (49.6%) belong to carbonyl chemistry, followed by substitution and elimination (26.8%), protection and deprotection (8.9%), reduction and oxidation (7.3%), functionalization and functional group conversion (6.8%), and a minor amount of olefination (0.6%). Each reaction superclass shows a different distribution of mechanistic steps and distinct chemical reactivity, highlighting the high diversity of the labeled reactions in this dataset. The reaction templates categorized to each reaction superclass can be found in Supplementary Information S2.

## Model architecture

To predict the electron flow by Reactron, the molecule structure is translated into molecule graphs, where the atoms and bonds in the given structures are represented as the nodes and edges in the graphs. We represent a set of input molecules, $M = (A, B)$, by a molecular graph, $G = (V, E)$, with vertices $(V)$ denoting heavy atoms ($A$) and edges ($E$) denoting bonds ($B$). Since the edge in the molecular graph is directional (i.e., $E_{uv}$ and $E_{vu}$ are served as different edges), the number of edges $|E|$ is twice of the number of bonds $|B|$ in the molecules. The molecular graph is built using DGL-LifeSci[23] and RDKit[24] python package, and the initial features of each vertex ($x_u^i$) and edge ($x_{uv}^i$) are described in Supplementary Information S3.

First, the local chemical environment of each heavy atom in the molecule is updated by gated-recurrent unit (GRU)-based message passing neural networks (MPNNs)[18] through aggregating the features from their neighboring atoms and bonds. We denote the message passing function by $MPNN(\cdot)$, with the atomic features of vertex $u$ as $x_u$, neighbor vertex features $\{x_v\}$ and neighbor edge features $\{x_{uv}\}$.

$$x_u^{t+1} = MPNN\left(x_u, \{x_v\}_{v\in V}, \{x_{uv}^i\}_{uv\in E}\right). \quad (1)$$

The hidden dimension of the GRU in MPNN is set to 128 and each node is updated for 3 iterations by message passing

After new features of each heavy atom ($x_a$) are acquired after the message passing, the features of the hydrogen attached to the heavy atom and the bond connected between atoms are calculated by a feature splitting module. In the feature splitting block, the features of the hydrogen $x_{u'}$ and bond $x_{uv}$ is made from the atom features $x_u$. The features of the hydrogen are first made by a non-linear transformation of the features of atom

$$x_{u'} = \sigma(W^h x_u + b^h) \quad (2)$$

, where $\sigma$ is the ReLU activation function. Note that since each hydrogen on the atom is assumed to be equivalent, only one hydrogen feature is made for each heavy atom. Thus, the molecular graph includes a total $2|V|$ vertices after including the hydrogen vertices.

Next, the bond features are made by concatenating the feature from each atom pair connected by the bond and a non-linear transformation

$$x_{uv} = \sigma(W^b(x_u||x_v) + b^b) \quad (3)$$

Due to the additional vertices from hydrogens, the total number of edges in the molecular graph increases from $2|B|$ to $2|A| + 2|B|$.

Then, the non-reactive atoms and bonds are identified and filtered out by the reactive pooling layers before considering all the potential reaction pathways between all atom and bond pair. This design aligns with the human chemist's intuition, where we only focus on a few well-recognized chemical species as the potential electron sources (nucleophilic sites such as the nitrogen atom in amines or the α-carbon in enol derivatives) and sinks (electrophilic sites such as carbonyl carbon or the α-carbon in alkyl halides) in the molecules when predicting the reaction path. In the reaction pooling layers, we train two classifiers to predict the scores of each atom or bond to be the source and sink. The atoms and bonds having the top-$k$ highest predicted source scores ($x_{source}$) and sink scores ($x_{sink}$) are selected to make the reaction path candidates, where the features of path candidates ($x_{path}$) are made by concatenating the features of sources, sinks and the one-hot encoding of their distance.

Two non-linear reactive pooling networks are trained to predict the probability of each atom or bond being the source and sink of the electron path.

$$\tilde{y}_a^{source} = W^{source2}\left(\sigma(W^{source1}x_a + b^{source1})\right) + b^{source2} \quad (4)$$

$$\tilde{y}_a^{sink} = W^{sink2}\left(\sigma\left(W^{sink1}x_a + b^{sink1}\right)\right) + b^{sink2} \quad (5)$$

The features of the potential electron paths are made by concatenating the features of all the source $a$ and sink $b$ pairs, together with their distances feature $r_{ab}$, resulting a total number of $k^2$ potential electron paths.

$$x_{ab} = x_a||x_b||r_{ab} \quad (6)$$

, where the distance feature $r_{ab}$ one-hot encoding of the distance $D_{a,b}$ between source $a$ and sink $b$

$$D_{a,b} = \begin{cases} distance(a,b), if\ distance\ (a,b) \leq 8 \\ 9, if\ disance(a,b) > 8\ and\ a\ and\ b\ are\ in\ the\ same\ molcule\,. \\ 10, if\ a\ and\ b\ are\ in\ difference\ molecules \end{cases} \quad (7)$$

Because the atom features and bond features are computed equivalent in the reactive pooling block, we use the symbol $a$ instead $u$ or $v$ of to represent each potential source and sink. According to the prediction scores, we preserve only the $k$ atoms or bonds with the top-k highest $\tilde{y}_a^{source}$ and $\tilde{y}_a^{sink}$ for further computation. The $k$ value is set to 8 in this study. With this reactive pooling, the number potential reaction paths are reduced from $O(V+E)^2$ to $O(k^2)$, where $V$ is the number of nodes (including heavy atoms and attached hydrogens) and $E$ is the number of edges (including bonds between heavy atoms and between heavy atoms and hydrogens, directional) in the molecules. As the average number of $V$ and $E$ in the dataset we use in this work are approximately 80 and 160, and the $k$ we used to be 8, the memory demand for the further computing is reduced by the order of 3 with reactive pooling.

Next, the reaction path candidates are sent to reactive attention blocks to exchange the features to facilitate the selectivity prediction. The reaction attention block is an attention mechanism applied in the molecule graph that was previously proposed in our previous models[8,14,25], showing great performance of enhancing the prediction for long-range atom/bond interactions.

The features of the pooled electron paths are updated by the reactivity attention to exchange the reactivity information through a multi-head attention mechanism based on a Transformer[26]-like architecture. The attention score between electron path $ab$ and $ac$ can be decomposed as $e_{ab,ac}$

$$e_{ab,ac} = \frac{x_{ab}W^Q(x_{ac}W^K)^T}{\sqrt{d_z}}, \quad (8)$$

The features are updated by a skipped connection and gated transformation same as LocalTransform[8]. This reactivity attention is repeated for 3 times for a sufficient information exchange. The default number of attention head is set to be 4 in our experiment.

Finally, the path score ($y_{path}$) is calculated by a non-linear classifier to decide the probability of each reaction path candidate. Note that the dimension of the last classification layer is $\mathrm{k}^2 + 1$, where an additional class representing the null path for training the Reactron to stop reaction when it meets the reaction product. The final path scores are normalized by a SoftMax function and the null path is excluded from the path candidates.

$$\tilde{y}_{ab}^{path} = W^{path2}\left(ReLU(W^{path1}x_{ab} + b^{path1})\right) + b^{path2} \quad (9)$$

$$s_{ab}^{path} = Softmax(\tilde{y}_{ab}^{path}) \quad (10)$$

## Training objectives

The total loss for training *Reactron* is the linear combination of the cross-entropy loss $CE(\cdot)$ of reactive pooling and path prediction with loss weight $\lambda$.

$$Loss = \lambda(CE(\tilde{y}_a^{source}, y_a^{source}) + CE(\tilde{y}_b^{sink}, y_b^{sink})) + CE(\tilde{y}_{ab}^{path}, y_{ab}^{path}) \quad (11)$$

The loss weight $\lambda$ is set to 1 at the first epoch and adjusted to 0.5 to the rest of the training process.

Since the electron movements in the chemical reaction could have multiple pathway other than the forming the major product, we apply label smoothing[27] to the loss of the path prediction with factor 0.01.

## Training hyperparameters

All the chemical operations are done using the RDKit[24] python package. We used Pytorch[28] and DGL for neural network training and testing. We train our model for 20 epochs with a batch size 16 by AdamW optimizer[29,30] with $10^{-6}$ weight decay and set the learning rate to $10^{-4}$.

## Implementation of baseline models

To assess the predictive performance of Reactron, we trained two product prediction model, Transformer[6] and Graph2SMILES[9], as the baseline models. For both product-only models and mechanism prediction Transformer, we follow the instructions given in the GitHub repository given in the original publications without any modification, including the model architecture and hyperparameters. For the mechanism models, we modify the codes of Graph2SMILES due to the invalidity of generating molecular graphs from the SMILES of the transition state molecules. In particular, we did not sanitize the molecule after converting the molecule SMILES into RDKit[24] molecule, and the number of valences and hydrogen was set to 6 and 4 if the number of valences and hydrogen could not be successfully calculated by RDKit.

## Finetuning Reactron and baseline models for named reactions

We finetuned Reactron and the baseline models to examine their ability to predict complex chemical transformation after learning from a few examples from the checkpoints saved after training on USPTO reactions. For Reactron, we changed the batch size from 16 to 128 and trained the model for 30 epochs. For Transformer and Graph2SMILES, we followed the finetuning strategy of Chemformer[31] to reduce the learning rate from 4 to 0.04 and the batch size from 4096 to 512, and increase the training epochs from 2,000 to 4,000. Because there is no stereo center in the training set of USPTO reactions, the baseline models could not recognize the unseen token of the stereo center (e.g. [C@] or [C@@]). Therefore, we eliminate the chiral information of each stereo center in the molecules, e.g. ignore the stereoisomer of the molecule SMILES, when finetuning and testing these baseline models.

## Data preparation for combinatorial reaction prediction

To prepare data from combinatorial reaction prediction, we extract the substrates and reagents of the reactions belonging to the top-400 reaction templates in the USPTO reaction dataset. For example, we can represent one reaction as $S + R >> P$, where $S$ is the substrate, $R$ is the reagent, and $P$ is the product of the reaction. By decomposing the substrate $S$ and reagent $R$ from all the reactions in the USPTO reaction dataset, a substrate pool $\{S\}$ containing $n$ unique substrates, and a reagent pool $\{R\}$ containing $m$ unique reagents. By pairing each substrate to each of the reagent in the reagent pool $\{R\}$, we can create a reaction matrix containing $n \times m$ reactions.

It is crucial to extract the proper substrate and reagents from the existing reactions to pair each reaction substrate and reagent to discover new reactivity. Although reagents are usually defined as the molecules that do not contribute any atom to the major product, such as DEAD and $PPH_3$ in the Mistunobu reaction, in the machine learning community during the data curation process[6,17,32–34], we note that several reactive reagents contribute their atoms in the major product, such as Grignard reagents or Wittig reagents. Therefore, simply extracting the molecules that do not contribute atoms to the major product cannot reflect the chemically recognized reagents. To solve this problem, we design a recursive algorithm to extract the important reagents from the reaction data. By assuming that reactions yielding the same reaction template share the same reagents, which is commonly used in the particular type of reactions, we recursively iterate the molecules in the reaction data to find the common reagents that are shared by at least $N$ reactions yielding the same reaction template, where $N$ is set to 30 in our experiment. The pseudocode of the algorithm is given in Supplementary Information S5.

By applying this algorithm to all the labeled reactions yielding the top 400 popular reaction templates, we extracted 250 unique reagent sets after removing the ones only containing common solvents such as toluene or acetonitrile. After defining the reagent pool, the substrates are defined by the molecules remained in the reactions after removing the reagents.

## Defining new transformation in reaction discovery

After predicting all reactions in the 3,000,000 reactions derived from the 100,000 reagent-template pairs, we analyze both the reaction templates extracted from the predicted reactions and the predicted electron movement. The reaction templates used in MechFinder[13] include the functional groups surrounding the reaction center for more detailed mechanism labeling. However, this sometimes leads to similar transformations being designated as different reaction templates due to variations in functional groups. To reduce the inclusion of similar transformations already present in the training set, we exclude any functional groups during the template extraction process, aligning with the local reaction templates (LRTs) used in LocalRetro[25]. If the extracted LRT exists within the LRT set derived from the training reactions, we classify the reaction as a "known transformation."

Furthermore, we observed cases where some reactions yield LRTs that are absent in the training set but are subsets of known reaction mechanisms. For example, the LRT of a [2+2] cycloaddition may not be explicitly found in the training set but should not be defined as a new transformation because it is part of a Wittig reaction mechanism learned by the model. Therefore, we also categorize a reaction as a "known transformation" if the predicted electron movement is found to be a subset of a labeled reaction mechanism in the training set.

Next, we apply a one-sided binomial test under the hypothesis that at least 50% of the reactions associated with the same template-reagent pair would yield the same valid reaction template. To select statistically significant prediction results, we retain only the template-reagent pairs for which at least 20 out of 30 reactions (p-value = 0.049) produce the same reaction template. For template-reagent pairs that do not exhibit significantly consistent predictions or yield a null reaction template, we categorize them as exhibiting no reactivity or inconsistent reactivity and exclude them from subsequent analysis. For template-reagent pairs that show consistent predictions and where more than 20 reactions exhibit known transformations, we categorize these pairs as "known transformations" and also exclude them from further analysis.

Only the template-reagent pairs that pass these two filtering steps are defined as representing "new transformations" in this study.

## References

1. Levy, D. E. *Arrow-Pushing in Organic Chemistry: An Easy Approach to Understanding Reaction Mechanisms*. (John Wiley & Sons, Inc., 2008).
2. Chen, J. H. & Baldi, P. No Electron Left Behind: A Rule-Based Expert System To Predict Chemical Reactions and Reaction Mechanisms. *J. Chem. Inf. Model.* **49**, 2034–2043 (2009).
3. Klucznik, T. *et al.* Computational prediction of complex cationic rearrangement outcomes. *Nature* **625**, 508–515 (2024).
4. Segler, M. H. S. & Waller, M. P. Neural-Symbolic Machine Learning for Retrosynthesis and Reaction Prediction. *Chemistry – A European Journal* **23**, 5966–5971 (2017).
5. Schwaller, P., Gaudin, T., Lányi, D., Bekas, C. & Laino, T. “Found in Translation”: predicting outcomes of complex organic chemistry reactions using neural sequence-to-sequence models. *Chemical Science* **9**, 6091–6098 (2018).
6. Schwaller, P. *et al.* Molecular Transformer: A Model for Uncertainty-Calibrated Chemical Reaction Prediction. *ACS Central Science* **5**, 1572–1583 (2019).
7. W. Coley, C. *et al.* A graph-convolutional neural network model for the prediction of chemical reactivity. *Chemical Science* **10**, 370–377 (2019).
8. Chen, S. & Jung, Y. A generalized-template-based graph neural network for accurate organic reactivity prediction. *Nat Mach Intell* **4**, 772–780 (2022).
9. Tu, Z. & Coley, C. W. Permutation Invariant Graph-to-Sequence Model for Template-Free Retrosynthesis and Reaction Prediction. *J. Chem. Inf. Model.* **62**, 3503–3513 (2022).
10. Lowe, D. M. Extraction of chemical structures and reactions from the literature. (2012) doi:10.17863/CAM.16293.
11. Bradshaw, J., Kusner, M. J., Paige, B., Segler, M. H. S. & Hernández-Lobato, J. M. A Generative Model For Electron Paths. Preprint at https://doi.org/10.48550/arXiv.1805.10970 (2019).
12. Joung, J. F. *et al.* Reproducing Reaction Mechanisms with Machine-Learning Models Trained on a Large-Scale Mechanistic Dataset. *Angewandte Chemie International Edition* **63**, e202411296 (2024).
13. Chen, S., Babazade, R., Kim, T., Han, S. & Jung, Y. A large-scale reaction dataset of mechanistic pathways of organic reactions. *Sci Data* **11**, 863 (2024).
14. Chen, S., An, S., Babazade, R. & Jung, Y. Precise atom-to-atom mapping for organic reactions via human-in-the-loop machine learning. *Nat Commun* **15**, 2250 (2024).
15. Coley, C. W., Green, W. H. & Jensen, K. F. RDChiral: An RDKit Wrapper for Handling Stereochemistry in Retrosynthetic Template Extraction and Application. *Journal of Chemical Information and Modeling* **59**, 2529–2537 (2019).

16. Chen, S. *et al.* Reaction Templates: Bridging Synthesis Knowledge and Artificial Intelligence. *Acc. Chem. Res.* **57**, 1964–1972 (2024).
17. Jin, W., Coley, C., Barzilay, R. & Jaakkola, T. Predicting Organic Reaction Outcomes with Weisfeiler-Lehman Network. *Advances in Neural Information Processing Systems* **30**, (2017).
18. Gilmer, J., Schoenholz, S. S., Riley, P. F., Vinyals, O. & Dahl, G. E. Neural Message Passing for Quantum Chemistry. Preprint at https://doi.org/10.48550/arXiv.1704.01212 (2017).
19. Kurti, L. & Czako, B. *Strategic Applications of Named Reactions in Organic Synthesis*. (Elsevier, 2005).
20. DeNinno, M. P., Etienne, J. B. & Duplantier, K. C. A method for the selective reduction of carbohydrate 4,6-*O*-benzylidene acetals. *Tetrahedron Letters* **36**, 669–672 (1995).
21. Cao, D., Xia, S., Li, L., Zeng, H. & Li, C.-J. PPh3-Promoted Direct Deoxygenation of Epoxides to Alkenes. *Org. Lett.* **26**, 6418–6423 (2024).
22. Denney, D. B., Vill, J. J. & Boskin, M. J. A New Synthesis of Cyclopropanecarboxylic Acids. *J. Am. Chem. Soc.* 84, 3944–3946 (1962).
23. Li, M. *et al.* DGL-LifeSci: An Open-Source Toolkit for Deep Learning on Graphs in Life Science. *ACS Omega* **6**, 27233–27238 (2021).
24. RDKit: Open-source cheminformatics. https://www.rdkit.org/.
25. Chen, S. & Jung, Y. Deep Retrosynthetic Reaction Prediction using Local Reactivity and Global Attention. *JACS Au* **1**, 1612–1620 (2021).
26. Vaswani, A. *et al.* Attention Is All You Need. Preprint at https://doi.org/10.48550/arXiv.1706.03762 (2023).
27. Szegedy, C., Vanhoucke, V., Ioffe, S., Shlens, J. & Wojna, Z. Rethinking the Inception Architecture for Computer Vision. Preprint at https://doi.org/10.48550/arXiv.1512.00567 (2015).
28. Paszke, A. *et al.* PyTorch: An Imperative Style, High-Performance Deep Learning Library. *Advances in Neural Information Processing Systems* **32**, 8026–8037 (2019).
29. Kingma, D. P. & Ba, J. Adam: A Method for Stochastic Optimization. *In 3rd Int. Conf. for Learning Representations* (2017).
30. Loshchilov, I. & Hutter, F. Decoupled Weight Decay Regularization. in *International Conference on Learning Representations* (2018).
31. Irwin, R., Dimitriadis, S., He, J. & Bjerrum, E. J. Chemformer: a pre-trained transformer for computational chemistry. *Mach. Learn.: Sci. Technol.* **3**, 015022 (2022).
32. Wigh, D. S., Arrowsmith, J., Pomberger, A., Felton, K. C. & Lapkin, A. A. ORDerly: Data Sets and Benchmarks for Chemical Reaction Data. *J. Chem. Inf. Model.* **64**, 3790–3798 (2024).
33. Lee, C., Chen, S., Ong, K. T., Yeo, J. & Jung, Y. Noise Analysis and Data Refinement for Chemical Reactions from US Patents via Large Language Models. Preprint at https://doi.org/10.26434/chemrxiv-2024-1zl02 (2024).
34. Toniato, A., Schwaller, P., Cardinale, A., Geluykens, J. & Laino, T. Unassisted noise reduction of chemical reaction datasets. *Nat Mach Intell* **3**, 485–494 (2021).